\documentclass[referee]{raa}

\usepackage{graphicx,times}
\usepackage{natbib}
\usepackage{amssymb,amsmath}
\bibpunct{(}{)}{;}{a}{}{,}
\usepackage[pagebackref=false]{hyperref}

\begin{document}

\title{$N$-Body Dynamics as a Mechanism for Mean Motion Resonance Formation in the Post-Gas Era}

\volnopage{Vol.0 (20xx) No.0, 000--000}
\setcounter{page}{1}

\author{Kai Sun 
      \inst{1}
   \and Sheng Jin 
      \inst{1}\thanks{Corresponding Author, ID: https://orcid.org/0000-0002-9063-5987.}
   \and Dong-Hong Wu 
      \inst{1}
   \and Meijuan Wang 
      \inst{1}
      }

\institute{Department of Physics, Anhui Normal University, Wuhu, Anhui 241002, China; {\it jins@ahnu.edu.cn}\\
\vs\no
{\small Received 202x month day; accepted 202x month day}}

\abstract{Mean motion resonances (MMRs) are traditionally thought to form during the gas-rich phase of protoplanetary disks via disk-driven migration. Whether MMRs can form solely through collisional damping after disk dispersal has not been systematically investigated. To study MMR formation in the post-gas-disk era, we perform a large suite of $N$-body simulations, each containing 10,000 planetesimals around an M-dwarf star and evolving over 10 Myr. We compare two groups of simulations that differ only in their initial mass distributions. Each group consists of 10 runs with statistically identical initial planetesimal orbital distributions, generated using different random seeds. One group adopts a uniform planetesimal mass distribution, whereas the other features a bimodal distribution with massive embryos embedded in a swarm of smaller planetesimals. At 1 Myr, 40\% of systems (8/20) exhibit near-resonant period ratios, but this fraction drops to 25\% (5/20) by 10 Myr as higher-order commensurabilities dissolve. Resonant-angle tracking shows that all persistent systems have circulating resonant angles at 10 Myr, although one system undergoes a transient libration phase between about 2 and 7 Myr before returning to circulation. These results demonstrate that collisional merging can drive MMR formation even without a gas disk, although in this scenario near-resonant configurations are strongly favored over true resonances. Furthermore, this mechanism produces similar MMR fractions in both groups, suggesting that MMR formation is insensitive to the initial mass distribution, although the latter may modulate the formation pathway to a limited extent.
}

\authorrunning{K. Sun, S. Jin et al.}
\titlerunning{$N$-body MMRs Formation}

\maketitle

\section{Introduction}
\label{sec:intro}

Mean motion resonances (MMRs) represent one of the most 
prominent orbital ordering phenomena in planetary systems. 
The commensurabilities of multi-planetary systems, where the orbital periods of two 
bodies satisfy $P_{\mathrm{out}}/P_{\mathrm{in}} \approx p/q$ 
with small integers $p$ and $q$, are observed across diverse 
astronomical contexts: from the resonant chains of 
TRAPPIST-1 \citep{Wu2016, Delisle2017, Gillon2017}, to the Galilean satellites locked in 
Laplace resonance, and the 
classic Neptune--Pluto 3:2 resonance in our solar system. 
The ubiquity of MMRs suggests 
that resonance capture is a robust outcome of planetary 
system evolution, yet the physical mechanisms enabling this 
capture remain a subject of active debate 
\citep{Kley2012, Izidoro2017, Dai2024}.

The classical paradigm establishes a clear formation 
picture: planets embedded in protoplanetary disks lose 
orbital energy through gas damping and are captured into 
resonant states during migration \citep{Goldreich1980, 
Ward1986, Lee2002}. Type~I migration, driven by density 
waves excited by low-mass planets \citep{Ward1997}, 
and Type~II migration, where massive planets open gaps and 
drift with the viscous disk \citep{Lin1986, 
Ward1997, Kretke2012}, both provide 
continuous damping considered necessary for MMR formation 
\citep{Terquem2007, Kajtazi2023}. In this 
framework, gas disks not only drive orbital decay but also 
provide the ``cooling'' mechanism for systems to evolve 
from chaotic to ordered configurations. The characteristic 
timescale for gas-driven resonance formation is 
$\tau_{\mathrm{gas}} \sim 10^{5}$\,yr 
\citep{Lee2002}. However, this paradigm has 
clear applicability boundaries.

Protoplanetary disks typically disperse within 3--10\,Myr 
\citep{Haisch2001}, leaving 
planets and remaining planetesimals to evolve in gas-free 
environments. Observations of systems such as TRAPPIST-1 
display a complex resonant architecture with high-order 
resonances (8:5, 5:3) among the inner planets and low-order 
(3:2) resonances among the outer planets \citep{Agol2021}. 
While these resonances may have formed during the gas-rich 
phase, the presence of fragile high-order commensurabilities 
suggests that the system may have experienced dynamical 
processing after gas disk dispersal. Collisional damping or 
scattering during the post-gas phase could modify or even 
re-establish resonant configurations, motivating the 
question of whether such mechanisms can operate 
efficiently in gas-free environments. 
The presence of fragile high-order commensurabilities in systems such as 
TRAPPIST-1 suggests that dynamical processing may continue to shape orbital 
architectures after gas disk dispersal. Whether collisional damping 
alone can drive planetesimal systems into resonant or near-resonant 
configurations in the absence of a gas disk, however, remains an open 
question. Addressing this would help clarify whether observed near-resonant 
systems are necessarily relics of gas-driven migration, or whether they can 
instead emerge from $N$-body dynamics in the post-gas era.

In the immediate post-gas phase following disk dispersal, 
collisions replace gas interactions as the dominant dynamical 
process. 
When two planetesimals collide and merge, the post-collision velocity of the resulting body is the mass-weighted average of their pre-collision velocities.
As a result, the more massive body contributes more significantly to the orbital motion of the merged product. During this inelastic merging process, kinetic energy is irreversibly dissipated from the system. The process conserves total mass and linear momentum while redistributing angular momentum.
The associated energy loss can serve as an effective ``cooling'' mechanism, analogous to gas damping. 
The collisional timescale can be 
estimated as 
$\tau_{\mathrm{coll}} \sim (n\sigma v)^{-1} \sim 
10^{3}$--$10^{4}$\,yr for a population of 10,000
 planetesimals with an initial surface density $\Sigma \sim 10^{3}\,\mathrm{g\,cm^{-2}}$, 
consistent with the minimum-mass solar nebula 
\citep{Hayashi1981}. Here, $n$ is the number density, 
$\sigma$ the collisional cross-section enhanced by gravitational 
focusing \citep{Ormel2010}, and $v$ the relative 
velocity. The cumulative effect over $\sim$1\,Myr of early 
post-gas evolution may be sufficient to drive resonance 
formation.
Furthermore, the post-gas dynamics of planetary embryos and planetesimals involve rich physical mechanisms that can affect their orbital evolution and lead to diverse multi-planetary architectures \citep{Rasio1996, Wu2003, Ogihara2009, Jin2011, Childs2025}.
 In this context, the observed resonant configurations of exoplanet systems are the combined result of gas-driven migration and post-gas dynamics.
 
Although collisions are ubiquitous 
in late-stage planet formation \citep{Leinhardt2005}, 
their effectiveness as a resonance formation mechanism 
in the immediate aftermath of disk dispersal still 
needs to be systematically validated 
\citep{Morrison2020, Childs2025}.
 Recent work on oligarchic growth has shown that resonant 
chains can survive the gas-free epoch only if systems 
avoid subsequent giant impacts and eccentricity excitation 
that would dislodge planets from resonance, suggesting 
that collisional damping could, in principle, facilitate 
orbital ordering during this critical transition 
\citep{Morrison2020}.
Recent GPU-accelerated $N$-body simulations demonstrate 
that collision-driven evolution alone can effectively 
establish resonant planetary configurations in gas-free 
environments \citep{Grimm2026}.

In this work, we focus on the effect of the initial mass distribution of planetesimals and embryos on resonance capture during the early gas-free phase. 
Within the violent, collision-driven $N$-body evolution that follows gas dissipation, the initial mass distribution emerges as an important modulating factor for subsequent orbital evolution.
In systems composed of low-mass, nearly equal-mass planetesimals, collisions are frequent and low-energy, providing ``gentle" damping.
In contrast, bimodal systems, where massive embryos are embedded in a background of small planetesimals, undergo high-energy mergers accompanied by ``violent'' energy dissipation.
To address how these different collisional damping mechanisms affect subsequent orbital evolution and whether they can leave traces on the final orbital configuration of the formed planetary system, we set up two comparison groups of $N$-body simulations with two specific objectives: (1) to verify whether collision damping alone---without gas-driven migration---can form and maintain MMRs configurations; and (2) to quantify the effect of initial mass distributions (uniform vs. bimodal) on the formation of MMRs within this early gas-free phase.

The remainder of this paper is organized as follows. 
In Section~2, we describe the $N$-body simulation setup, the initial conditions for the uniform-mass (UNI) and bimodal-mass (BI) planetesimal systems, and the resonance identification criteria, including both the period-ratio criterion and resonant-angle tracking. 
Section~3 presents the main results: the final orbital architectures at $t=10$\,Myr (Section~3.1), the evolutionary tracks of two representative simulations from the UNI and BI groups
 (Section~3.2), the resonance evolution histories based on the period-ratio criterion (Section~3.3), the ruling out of  true dynamical resonance through resonant-angle tracking (Section~3.4), and the systematic effects of the initial mass distribution on resonance formation (Section~3.5). 
In Section~4, we discuss the implications of our findings for observed post-gas systems, compare the collision-driven mechanism with classical gas-driven scenarios, and address the limitations of our model. 
Finally, we summarize our conclusions in Section~5.

\section{Methods}
\label{sec:methods}

We use the GPU-accelerated $N$-body code GENGA \citep{Grimm2014} to simulate the 
post-gas dynamical evolution of planetesimal systems, focusing 
on how the initial mass distribution modulates the efficiency 
of collision-driven MMRs formation.

\subsection{Simulation Setup and Initial Conditions}
\label{sec:setup}

 Our simulations model a planetary system around an M-dwarf star with stellar mass $M_\ast = 0.0898\,M_\odot$, similar to TRAPPIST-1 \citep{Gillon2018, Agol2021}. The total planetesimal mass is set to $M_{\rm tot} = 10\,M_\oplus$. 

The initial planetesimal annulus of $0.01$--$0.1$~AU is motivated by the observed orbital architecture of the TRAPPIST-1 system, whose planets reside at $0.011$--$0.063$~AU \citep{Agol2021}. This compact range allows us to test whether collisional processing in the immediate post-gas phase can produce resonant-like orbital configurations without invoking large-scale radial migration from several AU. We acknowledge that this initial range is considerably more confined than the terrestrial-planet formation simulations of \citet{Pan2022}, which typically employ initial embryos at $0.5$--$5$~AU. This difference reflects the observed characteristics of M-dwarf planetary systems: TRAPPIST-1 planets all orbit within $0.01$--$0.06$~AU, and the inner edge of its primordial disk may have been located near $0.01$--$0.02$~AU. Our setup is therefore tailored to the post-gas dynamical evolution of compact M-dwarf systems rather than Solar System-like wide disks. The short dynamical timescale in this compact region ($\sim$$10^{-2}$~yr at $0.01$~AU) further enables us to capture $\sim$$10^2$--$10^3$ collisional timescales within the first 1~Myr, making the computational cost tractable while still resolving the critical early post-gas evolution.

The $N$-body system is integrated over a period of 10~Myr. 
The first 1~Myr covers approximately 100--1000 collisional timescales for the initial swarm, 
allowing sufficient dynamical processing for the system to assemble planetary embryos and develop quasi-steady orbital configurations. 
The integration is extended to 10~Myr to test the long-term stability of these configurations after collisional damping subsides. 
The time step is fixed at $\Delta t = 0.5$~day. 
At each detected collision, the step is temporarily reduced to 
resolve the merger at the correct orbital phase.

The initial planetesimal swarm comprises $N = 10,000$ bodies 
with semimajor axes drawn from a power-law distribution 
of $\mathrm{d}N/\mathrm{d}a \propto a^{-1/2}$ over $a \in [0.01, 0.1]$~AU, 
corresponding to a disk surface density profile $\Sigma \propto r^{-3/2}$. Initial eccentricities 
$e \in [0, 0.02]$ and inclinations $i \in [0^\circ, 1^\circ]$ are 
uniformly randomized, representing a dynamically cold disk 
consistent with late-stage planet formation \citep{Pan2022}.
The remaining initial orbital angles are randomly assigned values between 
 $0^\circ$ to $360^\circ$.
In all the $N$-body integrations, all collisions are treated as 
perfectly inelastic merging, with no mass loss or fragmentation. 
This simplification is justified for the late accretion stage 
where collision velocities remain below the mutual escape 
velocity \citep{Leinhardt2012}.

To investigate the effects of different mass distributions on planet formation and MMRs formation, we set up two comparison groups. 
The UNI group consists of $N = 10{,}000$ near-equal-mass planetesimals with mean mass $\bar{m}_{\rm uni} = 0.001\,M_\oplus$. The mass of each body is drawn from a random distribution with $\pm 10\%$ around $\bar{m}_{\rm uni}$. 
This setup assumes an idealized scenario in which all planetesimals are roughly the same size, with a small standard deviation in the mass distribution.
The BI group contains $N_{\rm big} = 500$ large embryos with mean mass $\bar{m}_{\rm big} = 0.01\,M_\oplus$ and $N_{\rm small} = 9\,500$ small planetesimals with mean $\bar{m}_{\rm small} \approx 0.00053\,M_\oplus$. The large and small populations each comprise 50\% of the total mass, and the mass of each body is assigned with random $\pm 10\%$ variations around its respective mean.
 This distribution mimics the outcome of runaway growth while the gas disk is still present \citep{Kokubo1996}. 
To isolate the effect of the initial mass distribution from other factors, all other parameters are kept identical between the two groups. Both groups share the same total mass, total number of bodies, semi-major axis distribution, and orbital parameters. Ten simulations were performed for each group. 
Table~\ref{tab:mass_distribution} lists the initial mass distributions for both groups.

\begin{table}[!htbp]
\centering
\caption{Initial Mass Distributions for the UNI and BI Groups.}
\label{tab:mass_distribution}
\begin{tabular}{lcc}
\hline\hline
\noalign{\smallskip}
 & UNI & BI \\
\noalign{\smallskip}
\hline
\noalign{\smallskip}
\multicolumn{3}{l}{\textit{Number of Bodies:}} \\
\quad $\bar{m}_{\rm big} = 0.01\,M_\oplus$ & --- & 500 \\
\quad $\bar{m}_{\rm small} \sim 0.00053\,M_\oplus$ & --- & 9,500 \\
\quad $\bar{m}_{\rm uni} = 0.001\,M_\oplus$ & 10,000 & --- \\
\hline
\noalign{\smallskip}
\multicolumn{3}{l}{\textit{Mass Distributions:}} \\
Total mass ($M_\oplus$) & 10 & 10 \\
Scattering around $\bar{m}$ & $\pm$10\% & $\pm$10\% \\
\noalign{\smallskip}
\hline
\noalign{\smallskip}
\multicolumn{3}{p{0.9\linewidth}}{\small Note. Both groups share identical semi-major axis distribution 
($a \in [0.01, 0.1]$~AU, $\Sigma \propto r^{-3/2}$), initial eccentricities 
($e \in [0, 0.02]$), and inclinations ($i \in [0^\circ, 1^\circ]$). }
\end{tabular}
\end{table}

\subsection{Resonance Identification Criteria}
\label{sec:resonance}

We identify candidate mean-motion resonances between any two bodies using two complementary criteria: a period-ratio filter and resonant-angle tracking.

\subsubsection{Period-Ratio Criterion}
\label{sec:period_ratio}

For planets on inner and outer orbits with periods $P_{\rm out}$ and $P_{\rm in}$, a $p$:$q$ commensurability is considered present if
\begin{equation}
    \left|\frac{P_{\rm out}}{P_{\rm in}} - \frac{p}{q}\right| \bigg/ 
    \frac{p}{q} \leq 0.01.
    \label{eq:resonance}
\end{equation}
The tolerance of 0.01 is chosen to balance 
sensitivity and robustness, consistent with the convention commonly
adopted in population studies of near-resonant exoplanetary 
systems \citep{Fabrycky2014, Izidoro2017}. 
This threshold effectively identifies near-resonant configurations while filtering out coincidental period ratios that arise by chance. We detect a range of commensurabilities, including the 2:1, 3:2, 4:3, 5:3, 8:5, 5:2, 7:3, 3:1, 4:1, and 5:4 resonances. These span common resonances from low-order ($|p-q|\leq2$; e.g., 2:1, 3:2, 5:3, 3:1) to moderate-order ($|p-q|=3$; e.g., 5:2, 4:1, 8:5) and high-order ($|p-q|\geq4$; e.g., 7:3).

It should be emphasized that the period-ratio criterion alone identifies configurations based solely on period ratios approaching simple rational numbers, without distinguishing true dynamical resonance locking from stochastic clustering near commensurabilities.

\subsubsection{Resonant-Angle Tracking}
\label{sec:resonance_angle}

To verify whether a candidate $p$:$q$ commensurability represents genuine dynamical resonance locking, we compute the resonant angle
\begin{equation}
    \phi_{p:q} = p\lambda_{\rm out} - q\lambda_{\rm in} - (p-q)\varpi_{\rm in},
    \label{eq:resonance_angle}
\end{equation}
where $\lambda$ is the mean longitude and $\varpi$ is the longitude of pericentre. 
For a true mean-motion resonance, $\phi_{p:q}$ must librate (oscillate) around a fixed equilibrium value with bounded amplitude. 
In contrast, circulation---continuous monotonic evolution of $\phi_{p:q}$ through the full $[0^\circ, 360^\circ]$ range---indicates a merely near-resonant configuration in which the planets are not dynamically locked.

The orbital elements required to evaluate Equation~(\ref{eq:resonance_angle}) are directly output by GENGA at each snapshot. 
For each pair of bodies satisfying the period-ratio criterion (Equation~\ref{eq:resonance}), we track $\phi_{p:q}$ throughout the 10~Myr integration. 
A system is classified as being in true resonance only if both (i) the period-ratio condition is satisfied and (ii) the corresponding resonant angle exhibits clear libration over a sustained interval ($\gtrsim 10^4$~yr). If the period-ratio condition is met but the resonant angle circulates, the pair is classified as a near-resonant configuration. A system that exhibits such libration for a sustained interval but later escapes into circulation is classified as a transient resonant episode rather than a long-term true resonance.

\section{Results}
\label{sec:results}

\subsection{Overview of Simulation Results}
\label{sec:final_arch}

We now present the statistical outcomes of our 20 $N$-body simulations 
at $t = 10$~Myr. We examine the final orbital architectures, the number of planets formed, their masses and eccentricity distributions, and how different  
initial mass distributions influence the evolution and final orbital configuration of the resulting planetary systems.

\begin{figure}[!htbp]
\centering
\includegraphics[width=0.95\textwidth]{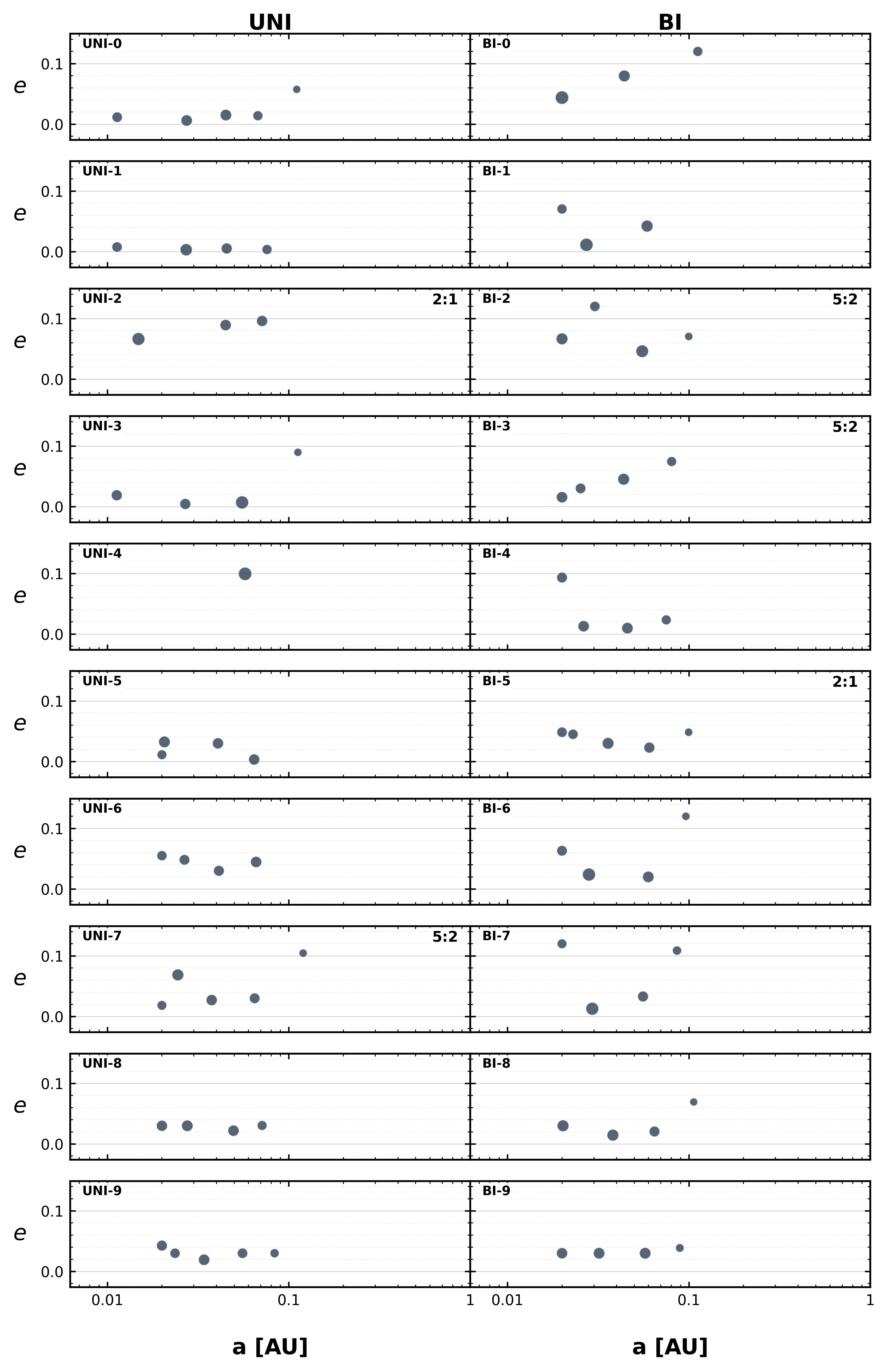}
\caption{ Final snapshots of semi-major axis versus eccentricity for all 20 simulations at 10~Myr. The left column shows the UNI group, and the right column shows the BI group. Point sizes are proportional to planetary mass. Upper-right labels indicate the near-resonant period ratios identified in that simulation at 10~Myr.}
\label{fig:final_10myr}
\end{figure}

 Figure~\ref{fig:final_10myr} presents the final distributions of surviving 
bodies in the semi-major axis versus eccentricity plane at 
$t = 10$~Myr for all 20 simulation groups. The left column shows the 
10 simulations from the UNI group, and the right column shows the 10 
simulations in the BI group. In each panel, the circle size scales with the mass of each body. The label in the upper-right corner indicates the 
near-resonant period-ratio type identified in that system at $t = 10$~Myr (e.g., 2:1, 5:2). 
Panels without an upper-right label denote simulations in which no period-ratio commensurability is identified.

Both the UNI and BI groups successfully form planets after 
10~Myr of collisional evolution, evolving into multi-planet systems 
regardless of initial mass distribution.  The final planet counts range from $N = 1$ 
to $N = 5$ across all systems at 10~Myr, reflecting continued collisional merging and dynamical scattering over the extended integration.
The simulations reveal a general trend in both groups: at 10~Myr, massive planets that have formed typically maintain low eccentricities in inner orbits ($e < 0.1$), while smaller planets in outer orbits exhibit higher eccentricities. These orbital architectures are consistent with typical terrestrial exoplanet systems.

 Near-resonant period-ratio configurations persist in 5 of 20 simulations (25\%) at 10 Myr: 2 in the UNI group (UNI-2 and UNI-7) and 3 in the BI group (BI-2, BI-3, and BI-5). This represents a decline from the 40\% (8/20) identified at 1 Myr, during which 3 systems lost their near-resonant status over the subsequent 9 Myr. This decline indicates that higher-order commensurabilities dissolve once collisional damping subsides, leaving only the lower-order and moderately low-order period ratios (2:1 and 5:2) in the final snapshot. As confirmed by resonant-angle tracking (Section \ref{sec:resonance_angle}), all persistent systems occupy near-resonant configurations rather than true dynamical resonances at 10~Myr, although the UNI-2 system does briefly sustain a true resonance before it dissolves. Taken together, these results suggest that, in the absence of gas-driven migration, collisional damping alone establishes near-resonant states with only limited efficiency---an efficiency that diminishes further over long-term evolution.

\subsection{Dynamical Evolution}
\label{sec:evolution}

To illustrate how collisional damping drives near-resonant orbital ordering during 
the post-gas phase, we examine the evolutionary history of two
representative systems from the BI and UNI groups that retain 2:1 period-ratio commensurabilities at 10~Myr.

\begin{figure}[htbp]
\centering
\includegraphics[width=\linewidth,keepaspectratio]{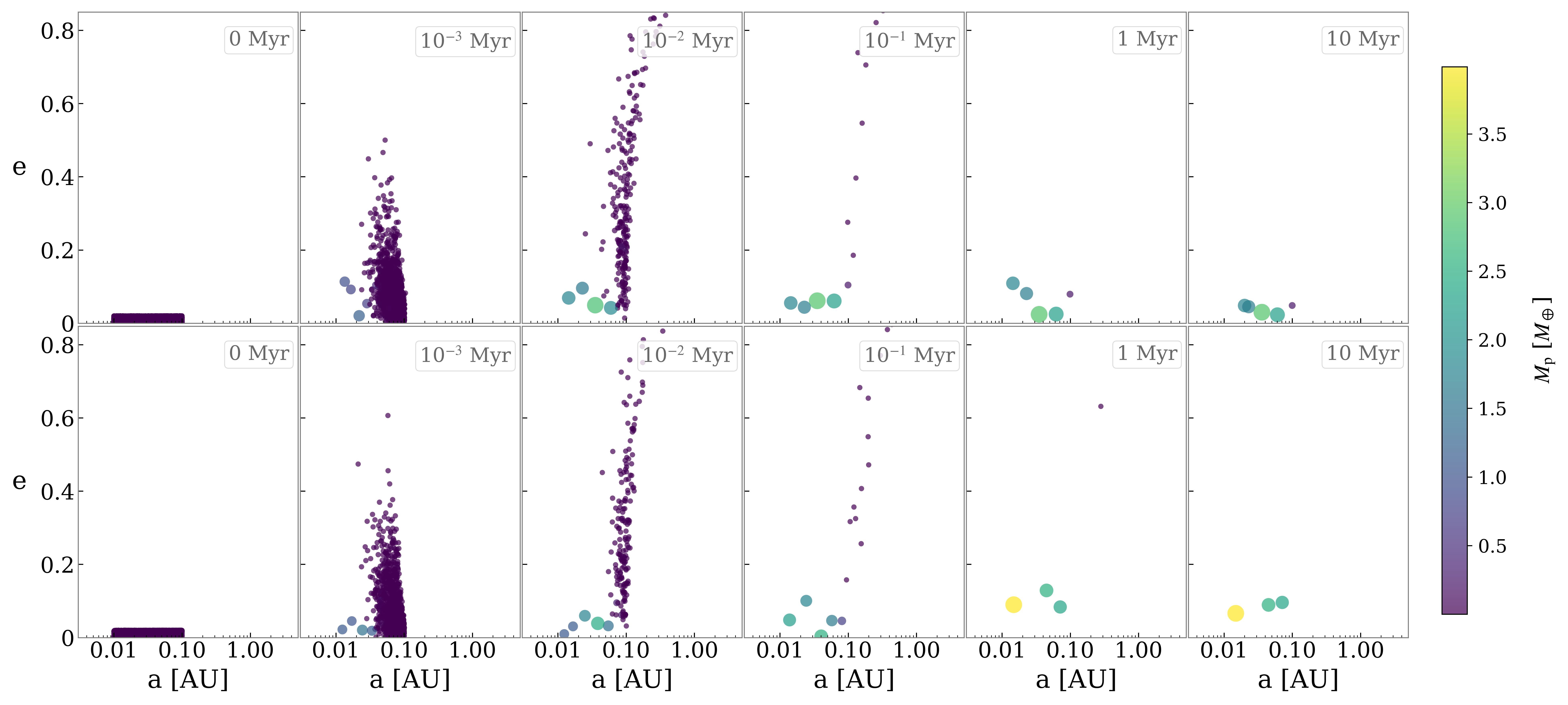}
\caption{ Evolutionary sequences of two representative systems that retain near-resonant period ratios at 10~Myr: BI-5 (upper panels) and UNI-2 (lower panels). Color denotes particle mass. Both systems experience strong dynamical excitation within the first $10^{-2}$\,Myr, followed by gradual eccentricity damping and finally converge to compact configurations with persistent 2:1 period-ratio commensurabilities at 10~Myr.}
\label{fig:evolution}
\end{figure}

 Figure~\ref{fig:evolution} presents the time evolution of two representative systems: BI-5 (upper panels) and UNI-2 (lower panels). The six panels for each system display the distribution of surviving planetesimals in the semi-major axis versus eccentricity plane at 0, $10^{-3}$, $10^{-2}$, $10^{-1}$, 1, and 10~Myr.

Both systems begin as dynamically cold disks with low eccentricities. 
As collisions commence, eccentricities are excited in both groups. 
By $10^{-3}$~Myr, bodies in both systems are dynamically heated to 
$e \sim 0.4$. At $10^{-2}$~Myr, both exhibit extended vertical spreads, 
with individual bodies reaching $e \sim 0.8$, while the number of 
surviving bodies decreases rapidly through mergers. By $10^{-1}$~Myr, most  planetesimals with large eccentricities have either been scattered out of the system or ended in a collisional merger. Only a small number of bodies remain in each system, as the systems have settled into a few larger planets.
From 1~Myr to 10~Myr, the systems continue to evolve with minimal collisional activity; the surviving planets maintain their compact orbital configurations, and the 2:1 period-ratio commensurability persists in both systems, though resonant-angle tracking (Section~\ref{sec:resonance_angle}) reveals that these are all near-resonant configurations  at 10 Myr. One exception is UNI-2, whose resonant angles librate over a brief interval spanning roughly 2--7 Myr before returning to circulation, pointing to a short-lived locking episode rather than permanent capture into resonance.
The similar evolutionary trajectories 
demonstrate that collisional damping provides an effective pathway 
to near-resonant configuration formation, regardless of whether the initial mass 
distribution is uniform or bimodal.

\subsection{Resonance Evolution}
\label{sec:resonance_evol}

To systematically characterize the near-resonant configuration formation process 
across all simulations that retained period-ratio commensurabilities, we examine the eight systems 
(4 UNI and 4 BI) that satisfy the period-ratio criterion at 
$t = 1$~Myr, and track their evolution out to 10~Myr. 
We track the evolution of the period-ratio commensurabilities in these systems from 0.3~Myr 
to 10~Myr at 0.1~Myr intervals, recording all candidate resonant pairs 
identified by the period ratio criterion (Eq.~\ref{eq:resonance}).
The results are summarized in Table~\ref{tab:resonance_history}. 
These temporal resolution reveals the rapid emergence of near-resonant architectures  during the early  phase, followed by the subsequent stabilization or disruption of these period-ratio pairs over longer timescales.
It provides a basis 
for comparing the evolutionary pathways between BI and UNI 
groups.

 By 10~Myr, only 5 of the original 8 systems retain period-ratio commensurabilities. In the UNI group, UNI-1 (7:3) and UNI-4 (7:3) lose their commensurabilities entirely, while UNI-2 maintains a 2:1 ratio and UNI-7 maintains a 5:2 ratio. In the BI group, BI-8 loses both its 7:3 and 2:1 commensurabilities, while BI-2 (5:2), BI-3 (5:2, having transitioned from the 5:3 commensurability present at 1~Myr), and BI-5 (2:1) retain their period ratios. The dissolution of the high-order 7:3, the moderate-order 8:5, and even the low-order 5:3 ratio—alongside the persistence of the lower-order 2:1 and moderately low-order 5:2 ratios—indicates that the width and dynamical stability of a commensurability play a critical role in its long-term survival. Theoretically, once collisional damping subsides, the absence of sustained dissipation can no longer maintain phase-coherent libration; higher-order resonances, intrinsically narrow and dynamically fragile, gradually diffuse and dissolve under long-term perturbations.

The UNI systems display relatively more complex evolutionary histories, characterized by intermittent period-ratio formation.
For example, UNI-2 hosts four commensurable pairs at 0.3~Myr, collapses to a single 5:3 at 0.4~Myr, loses all commensurabilities at 0.5--0.6~Myr, displays a transient 3:1 ratio at 0.7~Myr, and finally re-establishes the 2:1 commensurability from 0.8--1~Myr. Resonant-angle tracking further reveals that the 2:1 configuration in UNI-2 constitutes a genuine dynamical resonance from approximately 2 to 7 Myr. After this period, the system returns to circulation and persists only in a near-resonant state until 10 Myr. 
UNI-7 shows delayed formation, with its first 
commensurable pair (5:2) appearing only at 0.6~Myr,  and maintains this ratio through 10~Myr. At 1~Myr, each of the four UNI simulations retains exactly one period-ratio pair, and no system 
hosts multiple commensurabilities;  by 10~Myr, only two UNI systems retain near-resonant configurations. 

The BI systems exhibit relatively consistent period-ratio evolution. 
As the systems evolve, they rapidly converge toward 
simpler configurations. 
For instance, BI-3 transitions from multiple commensurable pairs 
at 0.5~Myr to a single stable 5:3 pair by 0.7~Myr, which persists 
until 1~Myr,  and ultimately settles into a 5:2 ratio by 10~Myr. 
 This transition from a 5:3 commensurability at 1~Myr to a 5:2 ratio 
at 10~Myr demonstrates that near-resonant configurations remain 
dynamically labile and can continue to shift long after the initial 
collisional damping subsides.
At 1 Myr, the four BI systems retain a total of 
six period-ratio pairs, with 50\% (2 out of 4) hosting multiple commensurabilities: BI-2 maintains two distinct 
5:2 pairs, and BI-8 hosts both a 7:3 and a 2:1 ratio.  By 10~Myr, BI-8 loses all commensurabilities, while BI-2, BI-3, and BI-5 each retain a single period-ratio pair.

Table~\ref{tab:resonance_history} lists the 8 out of 20 runs that formed near-resonant configurations at $t = 1$~Myr and traces their subsequent evolution until $t = 10$~Myr, at which point only 5 runs retain near-resonant configurations.
The BI systems tend to establish multiple commensurable ratios 
early and rapidly converge toward stable, low-order period-ratio configurations, owing to the strong gravitational influence of large planetary embryos in these systems. 
The UNI systems display more gradual evolutions, characterized by intermittent 
period-ratio loss and delayed commensurability establishment. 
Across all 8 runs at 1~Myr and 5 surviving runs at 10~Myr, these divergent pathways 
demonstrate that collisional damping---operating through perfectly 
inelastic mergers that irreversibly dissipates kinetic energy 
\citep{Leinhardt2012, Pan2022}---can effectively drive planetesimal systems 
into near-resonant period-ratio configurations regardless of whether the initial mass 
distribution is uniform or bimodal.
However, the timescale of dynamical processing and the final period-ratio multiplicity vary with the initial mass distribution,  and the long-term survival fraction drops from 40\% at 1~Myr to 25\% at 10~Myr.

\begin{table*}[htbp]
\centering
\scriptsize
\setlength{\tabcolsep}{8pt}
\caption{Resonance Evolution in Eight Systems (0.3--10~Myr)}
\label{tab:resonance_history}
\begin{tabular}{c|cccc|cccc}
\hline\hline
$t$ (Myr) & UNI-1 & UNI-2 & UNI-4 & UNI-7 & BI-2 & BI-3 & BI-5 & BI-8 \\
\hline
0.3 & 7:3, 8:5 & 2:1, 2:1, 5:3, 5:3 & 7:3, 5:3, 2:1 & -- & 5:2, 5:4 & 7:3, 7:3, 8:5, 5:3 & 2:1 & 7:3 \\
0.4 & 7:3, 7:3 & 5:3 & 2:1, 3:1, 7:3, 3:2 & -- & 5:2 & 7:3, 5:2, 5:2, 7:3, 5:3 & 2:1, 7:3 & 7:3, 2:1 \\
0.5 & 7:3 & -- & 2:1 & -- & 5:2, 3:2 & 7:3, 8:5, 5:2 & 7:3, 2:1 & 2:1 \\
0.6 & 7:3 & -- & 2:1 & 5:2 & 5:2 & 5:3, 5:3, 7:3 & 7:3, 2:1 & 2:1 \\
0.7 & 7:3 & 3:1 & 7:3, 2:1 & 5:2 & 5:2 & 5:3 & 2:1 & 7:3, 2:1 \\
0.8 & 7:3 & 2:1 & 7:3, 2:1 & 5:2 & 5:2 & 5:3 & 2:1 & 7:3, 2:1 \\
0.9 & 7:3 & 2:1 & 7:3, 2:1 & 5:2 & 5:2 & 5:3 & 2:1 & 7:3, 2:1 \\
1 & 7:3 & 2:1 & 7:3 & 5:2 & 5:2, 5:2 & 5:3 & 2:1 & 7:3, 2:1 \\
10 & -- & 2:1 & -- & 5:2 & 5:2 & 5:2 & 2:1 & -- \\
\hline
\end{tabular}
\end{table*}

\subsection{Resonant-Angle Tracking}
\label{sec:resonance_angle}

 To determine whether the persistent period-ratio commensurabilities at 10~Myr represent genuine dynamical resonance locking or merely near-resonant configurations, we compute the resonant angle $\phi_{p:q}$ (Equation~\ref{eq:resonance_angle}) for all five surviving systems (UNI-2, UNI-7, BI-2, BI-3, and BI-5) throughout the 10~Myr integration.

\begin{figure}[!htbp]
\centering
\includegraphics[width=\textwidth]{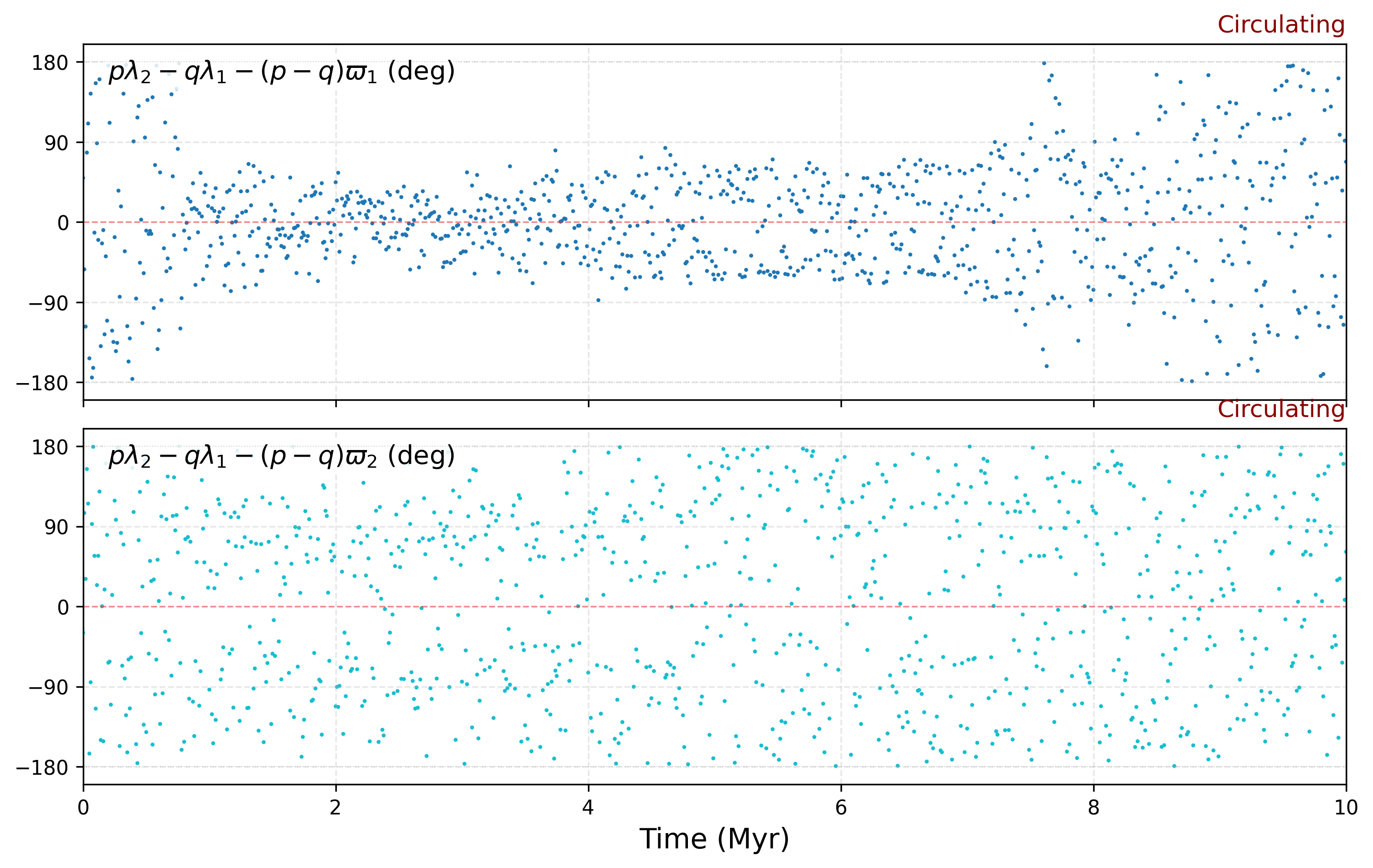}
\caption{Evolution of the resonant angle $\phi_{2:1}$ for the persistent period-ratio pair in UNI-2 from 0 to 10~Myr. 
As shown in the upper panel, the angle librates from approximately 2 to 7 Myr, indicating temporary phase locking, before escaping to circulation after about 7 Myr.
At 10~Myr, it circulates through the full $[0^\circ, 360^\circ]$ range, indicating a near-resonant configuration rather than true dynamical resonance locking. The other four persistent period-ratio pairs also exhibit circulation at 10~Myr.}
\label{fig:resonance_angle}
\end{figure}

Because four of the five persistent period-ratio pairs exhibit qualitatively similar circulation behaviour---continuous monotonic evolution of $\phi_{p:q}$ through the full $[0^\circ, 360^\circ]$ range without any bounded libration---while the remaining pair (UNI-2, 2:1) shows a transient libration phase, we show UNI-2 as a representative example in Figure~\ref{fig:resonance_angle}.

Figure~\ref{fig:resonance_angle} presents the evolution of the resonant angle for the 2:1 pair in UNI-2 from 0 to 10~Myr. The upper panel reveals that the 2:1 resonant angle remains in libration from approximately 2 to 7~Myr, indicating temporary phase locking. After about 7~Myr, however, the libration disappears and the angle escapes into continuous circulation through the full $[0^\circ, 360^\circ]$ range, so the system is not in true resonance at 10~Myr. The other four persistent period-ratio pairs---the 2:1 pair in BI-5 and the 5:2 pairs in UNI-7, BI-2, and BI-3---exhibit continuous circulation throughout the entire 0--10~Myr interval.

This demonstrates that every period-ratio commensurability identified in our simulations that persists from 1 to 10 Myr fails to qualify as true mean-motion resonance locking. Although UNI-2 exhibits a transient resonant episode between 2 and 7 Myr, the systems ultimately occupy near-resonant configurations in which the planets are not dynamically coupled by the resonance. The absence of sustained libration indicates that collisional damping alone, without the continuous dissipation provided by a gas disk, is insufficient to establish the phase-coherent dynamical capture required for genuine MMR formation \citep{Terquem2007, Marzari2018, Hu2025}. The evolution from 8 near-resonant systems at 1~Myr to only 5 at 10~Myr demonstrates that, in the absence of a phase-locking mechanism, these configurations do not possess the long-term protective mechanism characteristic of true mean-motion resonances. In the absence of additional late-stage dissipation---such as friction from a remnant planetesimal disk or stellar tides---these configurations are expected to become unstable and dissociate over longer timescales \citep{Hu2025}.

\subsection{Impact of Mass Distribution on Near-Resonant Formation}
\label{sec:mass_effect}

Comparing the final near-resonant states at 10~Myr reveals systematic differences between UNI and BI groups, as quantified in Table~\ref{tab:resonance_comparison}. The BI systems retain slightly more simulations with period-ratio commensurabilities than the UNI systems (3 out of 10 vs.\ 2 out of 10). Given the limited sample size (10 runs per group), this difference is not statistically significant and should be interpreted as a qualitative trend rather than a robust statistical result. Notably, all surviving commensurabilities at 10~Myr are low-order or moderately low-order (2:1 and 5:2); no moderate- or high-order ratios (e.g., 7:3, 8:5) persist in either group.

The long-term stability also differs. BI systems exhibit continuous resonant occupation from approximately 0.3--0.4~Myr onward, with three systems maintaining commensurabilities to 10~Myr. In contrast, the UNI systems generally exhibit more variable and turbulent resonance evolution. For example, UNI-2 experiences a complete loss of resonances at 0.5--0.6~Myr before re-establishing 2:1 at 0.8~Myr, which persists as a near-resonant configuration to 10~Myr, whereas UNI-7 exhibits delayed resonance formation, with its first resonance pair---5:2---appearing only at 0.6~Myr, and retaining it to 10~Myr.

 Although both the UNI and BI groups exhibit the same overall near-resonant occurrence rate at 1~Myr (40\%), the 10~Myr statistics reveal a divergence: BI systems retain a 30\% near-resonant fraction, while UNI systems drop to 20\%. This suggests that the presence of massive embryos in the BI systems may provide stronger gravitational stabilization for period-ratio commensurabilities, enhancing their long-term survival probability. However, resonant-angle tracking (Section~\ref{sec:resonance_angle}) confirms that even these persistent BI configurations are not true dynamical resonances. Given the limited sample size in this study (only 10 simulations per group, of which just 2 in UNI and 3 in BI formed near-resonant pairs at 10~Myr), these differences should be interpreted as indicative trends rather than statistically robust conclusions. Larger simulation ensembles, longer integration times, and a broader exploration of initial mass distributions will be necessary to confirm and generalize these findings.

\begin{table}[htbp]
\centering
\caption{ Resonance statistics at 10~Myr across all 10 simulations for UNI and BI groups}
\label{tab:resonance_comparison}
\begin{tabular}{@{}lcc@{}}
\hline\hline
\noalign{\smallskip}
 Fraction (out of 10 runs) & UNI  & BI  \\
\noalign{\smallskip}
\hline
\noalign{\smallskip}
Near-Resonant Systems (period ratio) & 20\% & 30\% \\
Low-Order Resonances ($|p-q|\le 2$)  & 10\%  & 10\%  \\
Multiple Resonance Pairs & 0\%  & 0\%  \\
Surviving Pair at 10~Myr & 20\% & 30\% \\
\noalign{\smallskip}
\hline
\end{tabular}
\end{table}

\section{Discussion}
\label{sec:discussion}

In this section, we discuss our simulation results within the broader context of planetary system formation. We first compare the collision-driven near-resonant formation scenario with observed exoplanet systems, exploring whether the predicted period-ratio architectures and formation probabilities are compatible with existing exoplanet data. We then contrast the collisional damping mechanism with classical gas-driven migration paradigms, highlighting both similarities and fundamental differences in their dynamical pathways. Finally, we discuss the limitations of our current model and identify directions for future work.

\subsection{Comparison with Exoplanet Systems}
\label{sec:obs}

Our simulations demonstrate that near-resonant orbital configurations can emerge via collisional damping in gas-free environments, for both  UNI and BI initial mass distributions. To connect this mechanism with observations, we consider the TRAPPIST-1 system as a qualitative benchmark. Although our single-annulus planetesimal model cannot fully reproduce the complex multi-planet architecture of the TRAPPIST-1 system, the observed diversity of resonance orders in this system---low- and moderate-order (5:3, 8:5) among inner planets and low-order (3:2) among outer planets \citep{Agol2021}---serves as a useful reference. Both our UNI and BI groups similarly produce a mixture of high-order and low-order period-ratio commensurabilities at 1~Myr. This similarity suggests that such diversity of near-commensurable pairs can arise during the gas-free $N$-body phase after disk dispersal.

The $\sim 25\%$ near-resonant occurrence rate identified by the period-ratio criterion at 10 Myr (5 out of 20 systems) provides a natural quantitative comparison with the observational constraints from \textit{Kepler} multi-planet systems. \citet{Izidoro2021} show that up to $\sim 95\%$ of resonant chains become dynamically unstable after gas disk dispersal, with only $\lesssim 5\%$ surviving as stable resonant configurations (e.g., the TRAPPIST-1 system). This implies that true dynamical resonance locking is intrinsically rare in exoplanet populations. Our results are consistent with this picture: collisional processing drives $\sim 25\%$ of systems into near-resonant period-ratio configurations by 10 Myr, yet resonant-angle tracking confirms that none of these correspond to true resonance locking, with the sole exception of UNI-2, which undergoes genuine resonant libration between approximately 2 and 7~Myr before the resonance ultimately dissolves. The fact that our near-resonance fraction ($25\%$) is substantially higher than the stable resonance fraction in observations ($\lesssim 5\%$) \citep{Izidoro2021} reflects the fundamental difference between near-resonant configurations (circulation, transient) and true resonances (libration, dynamically locked). The intermittent near-resonant histories in UNI systems (such as UNI-2, which loses all period-ratio commensurabilities at 0.5--0.6~Myr before re-establishing a 2:1 ratio at 0.8~Myr) further suggest that these near-resonant period ratios frequently represent transient evolutionary stages rather than final, stable configurations \citep{Hu2025}. Indeed, even the genuine 2:1 resonance that UNI-2 temporarily sustains between approximately 2 and 7~Myr ultimately dissolves, with its resonant angles returning to circulation, underscoring that even true resonances established through collisional processing can be short-lived in the post-gas-disk era.

Longer-term integration and resonant-angle tracking reveal that the near-resonant configurations identified at 10~Myr do not correspond to true dynamical resonance locking \citep{Goldberg2023}. The persistent 2:1 and 5:2 period ratios at 10~Myr exhibit resonant-angle circulation rather than libration, indicating that collisional damping alone---without continuous gas-driven dissipation \citep{Terquem2007, Kajtazi2023}---tends to produce transient near-resonant orbital orderings rather than long-term stable resonance capture.
Although UNI-2 temporarily maintains a genuine 2:1 resonance from roughly 2 to 7 Myr, the system ultimately decays into a near-resonant configuration by 10 Myr.
This offers a natural dynamical explanation for the observed excess of near-resonant systems and the corresponding deficit of strictly resonant systems in exoplanet populations: collisional processing can drive period ratios toward commensurabilities, but without sustained damping, the resonant phase relation cannot be maintained. 

Our simulation results provide a viable gas-free pathway for generating near-resonant architectures, and even---albeit with low probability---temporary true resonance configurations after gas dispersal. Unlike gas-driven migration, the collisional damping scenario does not require the presence of a gas disk, thus offering an alternative route to orbital commensurability formation. Similar conclusions have been reached by recent large-scale $N$-body simulations, which also demonstrate that collision-driven evolution can effectively produce near-resonant period-ratio pairs in the absence of gas disks \citep{Grimm2026}.

\subsection{Comparison with Gas-Driven Migration}
\label{sec:gas}

The classical paradigm for MMR formation posits that gas disks are necessary for true resonance capture, with Type~I and Type~II migration providing continuous damping \citep{Terquem2007, Dai2024}. In contrast, our simulations explore an alternative energy-dissipation channel. After gas disk dispersal, frequent $N$-body collisions efficiently dissipate kinetic energy on timescales $\tau_{\rm coll}\sim10^3$--$10^4$\,yr. The cumulative effect of this collisional damping appears sufficient to drive planets into near-resonant period ratios.

While gas damping acts continuously and uniformly, collisional damping is inherently discrete and intermittent. In the bimodal initial mass distribution (BI) systems, the strong gravitational influence of massive embryos leads to relatively stable near-resonant evolution during the first 1~Myr. In contrast, the uniform initial mass distribution (UNI) systems exhibit more variable and chaotic period-ratio histories. Nevertheless, both the BI and UNI systems ultimately evolve into similar near-resonant configurations at 10~Myr. This indicates that, in the post-gas-disk phase, collisional damping indeed provides an effective pathway for establishing orbital commensurabilities, and that this process is insensitive to the initial mass distribution of planetesimals.

However, the formation of true mean-motion resonances (MMRs) is generally believed to rely on the presence of a gas disk. In the classical theoretical framework, planetary migration in a gas disk---particularly Type~I and Type~II migration---can provide continuous and sustained orbital damping, allowing adjacent planets to be captured into mean-motion resonances during convergent migration, and maintain long-term libration of the resonant angles \citep{Terquem2007, Izidoro2021, Bitsch2024}.Without such sustained external dissipation, systems have difficulty establishing and maintaining the phase-coherent dynamical structures required for true resonance. 

This comparison highlights a fundamental physical distinction: gas-driven migration provides continuous, externally imposed damping that can effectively lock planets into true resonant configurations and maintain the resonant equilibrium against overstability-driven escape \citep{Terquem2007, Goldreich2014, Kajtazi2023}. In contrast, collisional damping is an internal, self-limiting process that dissipates kinetic energy only during discrete merger events; hence, the resulting near-resonant configuration may be a transitional architecture that is unstable over the long term.

Once the collisional cascade gradually weakens, the system transitions into a phase with sparse planetesimals and a dynamically ``quiet'' state. At this point, the resonant phase relationship is no longer actively maintained. The intermittent nature of collisional damping in our simulations indicates that establishing and sustaining the phase-coherent dynamical capture required for true MMRs is difficult in a collision-driven scenario. Even if a system briefly exhibits near-resonant period-ratio configurations in its early evolution, such states tend to be transitional rather than stable resonant lockings. Admittedly, it remains observationally unclear whether the commensurabilities observed in exoplanet systems are genuinely in resonant locking states, as measuring resonant angles requires high-precision transit-timing or radial-velocity data. Future higher-precision observations and more comprehensive $N$-body simulations will help to clarify this issue.

\subsection{Limitations and Future Work}
\label{sec:limit}

Several limitations of this study warrant discussion. First, 
our statistical sample comprises only 20 simulations (10 
per group). While the general trends---BI systems showing higher 
10~Myr survival fractions (30\% vs.\ 20\%), 
they are not yet statistically robust. Furthermore, although true dynamical resonances may form transiently during the evolution, the resonant-angle analysis confirms that none of the persistent period-ratio pairs at 10~Myr remain in true resonance, implying that it is extremely difficult to form and maintain true resonance in a collision-driven scenario. Future work will expand the simulation ensemble to fully explore these findings.

Second, our treatment of collisions as perfectly inelastic 
mergers with no mass loss or fragmentation  
is idealized. While this is reasonable for low-velocity encounters 
below the mutual escape speed, it breaks down for the high-velocity collisions prevalent in our simulations, which produce planets with eccentricities as high as $e\sim0.8$ prior to 
circularization---indicating high collisional velocities.
At such high velocities, hit-and-run collisions or 
catastrophic disruption may become significant, potentially altering the mass 
distribution and damping efficiency. Incorporating a 
fragmentation model would represent an important next step \citep{Leinhardt2012}.

Third, we have verified our period-ratio identifications through resonant-angle tracking (Section~\ref{sec:resonance_angle}). The 10~Myr analysis reveals that  all persistent period ratios---including the 2:1 and 5:2 configurations that survive from 1~Myr to 10~Myr---exhibit resonant-angle circulation rather than libration, indicating near-resonant rather than true dynamical resonance locking. However, one system (UNI-2) shows a transient libration phase between approximately 2 and 7~Myr before escaping back to circulation, indicating temporary phase locking rather than long-term resonance capture. Higher-order commensurabilities (e.g., 7:3) dissolve within a few Myr once collisional damping subsides. These findings suggest that collisional damping alone---without continuous gas-driven dissipation---tends to produce transient near-resonant configurations. Future work should examine the interplay between collisional damping and residual gas drag in a hybrid post-gas model, as well as explore a broader range of initial semi-major axes to test the robustness of our conclusions.

Fourth, our choice of a narrow initial annulus ($0.01$--$0.1$~AU) differs from the wider disks ($0.5$--$5$~AU) typically adopted in late-stage accretion models \citep{Pan2022}. While this confined setup brackets the current TRAPPIST-1 orbits and makes the critical early collisional evolution computationally tractable, planetesimals may indeed form at larger radii and migrate inward. A more extended disk would increase the local dynamical and collisional timescales by  orders of magnitude, potentially delaying planet assembly. However, the qualitative behaviour---collisional mergers dissipating kinetic energy and driving period ratios toward commensurabilities---would remain unchanged.

\section{Conclusions}
\label{sec:conclusions}

We have investigated the formation of near-resonant orbital configurations in the post-gas era through 
collisional damping alone, using a large suite of $N$-body simulations with 
two distinct initial mass distributions integrated to 10~Myr. Our main 
conclusions are as follows:

\begin{enumerate}
    \item Collisional damping via merging provides an effective mechanism for driving planetesimal systems into near-resonant orbital configurations in gas-free environments. Over the first 1~Myr of $N$-body evolution, $\sim$40\% of simulated systems (8 out of 20) establish near-resonant period ratios. However, extending the integration to 10~Myr reduces this fraction to 25\% (5 out of 20), as higher-order commensurabilities dissolve once collisional damping subsides. Furthermore, resonant-angle tracking reveals that at 10~Myr all persistent period ratios---including the 2:1 and 5:2 configurations that survive to 10~Myr---exhibit circulation rather than libration, indicating that these are near-resonant configurations rather than true dynamical resonance states. One system (UNI-2) nevertheless shows a transient libration phase between approximately 2 and 7~Myr before escaping back to circulation, indicating temporary phase locking rather than long-term resonance capture. These findings suggest that while collisional processing in the immediate aftermath of disk dispersal can serve as a complementary ``cooling'' mechanism for orbital ordering, it is difficult to establish or maintain true mean-motion resonance locking over Myr timescales. Sustained resonance locking likely requires the continuous dissipation provided by protoplanetary gas disks.
    
    \item  The initial mass distribution slightly modulates the near-resonant
    formation pathway and its long-term survival. Simulations in the BI
    group retain slightly more period-ratio commensurabilities to 10~Myr
    than the UNI group (3 out of 10 vs.\ 2 out of 10), Given the limited
    sample size (10 runs per group), this difference is not statistically significant and should not be over-interpreted as a qualitative trend. In general, both the UNI and BI groups form only near-resonance configurations, and the formation rate of these near-resonances are similar, although the BI group shows a little higher rate owing to one additional case.
    Taken together, these results suggest that the collision mechanism itself is intrinsically robust, regardless of the different mass distributions of the planetesimal swarm, although the strength of this relationship, and how the near-resonance formation rate depends on the initial mass distributions of planetesimals, still requires further validation with larger simulation ensembles.
    
    \item  Although our results reveal a possible connection between the characteristics of near-resonant statistics and the initial mass distributions after gas-disk dispersal---with the BI group generally favoring higher long-term survival of period-ratio commensurabilities---these correlations are subtle, since both BI and UNI groups produce a considerable number of near-resonant systems at 10~Myr. More importantly, resonant-angle analysis demonstrates that none of these systems, regardless of initial mass distribution, achieve long-term true dynamical resonance locking in the absence of continuous gas-driven damping. The UNI-2 system does exhibit a transient true resonance, but it ultimately dissolves. In summary, from the perspective of post-gas $N$-body dynamics, it is difficult to infer the planetesimal mass distribution based solely on near-resonant period-ratio statistics. Moreover, as resonant-angle libration—the definitive signature of true resonance—remains largely unmeasurable for most exoplanetary systems, it is still unknown whether the observed near-resonant pairs are genuinely locked or merely circulating near commensurabilities. Our results suggest that true resonance locking, if realized in nature, requires physical mechanisms beyond pure collisional damping, most plausibly the continuous dissipation provided by protoplanetary gas disks.
    
\end{enumerate}

{Future work that addresses the limitations of this study will enable a more robust exploration of our findings. First, we will expand the number of simulations and the range of initial mass distributions to enable a more quantitative assessment of the formation efficiency of near-resonant configurations, its dependence on the initial conditions of planetesimal distributions, and the low probability of true resonance.
Second, incorporating collisional fragmentation and hit-and-run interactions will clarify whether our idealized perfect-merger assumption significantly alters the damping efficiency or final near-resonant states. Third, extending the integration timescale beyond 10\,Myr will test the long-term stability of collision-driven resonances. Exploring a broader range of initial semi-major axes---from compact M-dwarf disks to wider Solar System-like configurations---will test the robustness of our conclusions across diverse planetary system architectures. Moreover, examining the interplay between collisional damping and residual gas drag in a hybrid post-gas model will bridge the gap between our purely collision-driven scenario and the classical gas-driven paradigm. Gas disks disperse gradually rather than instantaneously, and a more realistic model could reveal how gradual gas dissipation interacts with collisional damping to influence the transition from near-resonant orbital ordering to true resonance locking. Collectively, these extensions will help map the parameter space of collision-driven near-resonant formation and constrain its potential relevance to the observed diversity of exoplanetary systems.}

\begin{acknowledgements}
The authors are grateful to the referee for the constructive comments that have significantly improved this manuscript.
S.J. acknowledges the financial support from the National Natural Science Foundation of China (NSFC) under No.11973094 and the Incubation Program of Anhui Normal University (2023GFXK153). D.W. acknowledges the support from NSFC under No.12573076.
\end{acknowledgements}

\label{lastpage}

\end{document}